\documentclass[12pt]{article}
\usepackage{amssymb}
\usepackage{amstex}
\title{Electron and Phonon Thermal Waves in Semiconductors: an 
       Application to Photothermal Effects}
\author{G. Gonz\'{a}lez de la Cruz and Yu.~G.~Gurevich,\\
        Departamento de F\'{\i}sica,\\
Centro de Investigacion y de Estudios Avanzados del I.P.N.,\\
Apartado Postal 14--740, C.P. 07000,\\
M\'{e}xico, D.F., M\'{e}xico}

\begin{document}
\maketitle

\begin{abstract}
The electron and phonon temperature distribution function are calculated
in semiconductors.  We solved the coupled one-dimensional
heat-diffussion equations in the linear approximation in which the physical
parameters on the sample are independent of the temperature.  We also 
consider the heat flux at the surface of the semiconductor as a boundary
condition for each electron and phonon systems instead of using a fixed
temperature.  From this, we obtain an expression for electron and phonon
temperature respectively. The characterization of the thermal waves
properties is duscussed and some practical procedures for this purpose
provide us information about the electron and phonon thermal parameters.
\end{abstract}

\section{Introduction}
Thermal wave physics is becoming a valuable tool in the study of 
material parameters as well as in the semiconductor industry for
characterizing process in the manufacturing of electronic
devices.$^{1-2}$  These waves are created whenever there is a periodic
heat generation in a medium.  The most common mechanism for producing
thermal waves is the absorption of an intensity modulated light beam by
a sample.  The photothermal heating of a sample leads to thermal ant
stress-induced changes in the physical properties of both the sample and
the surrounding madia. Based on this observation, several alternative
detection techniques have been developed for monitoring the
photothermally induced changes of given physical properties either of
the sample or of its surrounding medium. Some of the different 
contact-type detection techniques, such as the conventional 
gas-microphone photoacoustic detection,${}^{1}$ photopyroelectric 
detection,${}^{3}$ or remote semsing techniques as photothermal 
reflection${}^{4}$ etc., have been reviewed in Ref.~5.

Being a photothermal techniques, the detected signal is strongly 
dependent upon how the heat diffuses through the sample allows us to
perform both thermal characterization of the sample (i.e., measurements
of its thermal properties, such as thermal diffusivity and thermal
conductivity), and carrier-transport properties.$^6$ In the case of
semiconductors the photoacoustical signal can provided us with
additional information regarding the heat-transport properties and
electron--phonon energy relaxation by the interacting electron and
phonon systems, a fact which has been recognized since the earlier
theory of heat conduction in semiconductors and hot electrons$^{6--8}$
Qualitatively, this may be understood as follows: in studies of the
behavior of semiconductors heating by electromagnetic waves it is usual
to assume that the phonon system remains in equilibrium, i.e., that the
phonon temperature $T_p$ is equal to the temperature $T_0$ of the
ambient medium.${}^{8}$ However, nonequilibrium carriers in the
bulk of the sample can interact and transfer energy to the phonon 
system producing an essential increase in the average energy of phonons
(energy nonequilibrium), which can be most conveniently described in
terms of heating of the phonon gas (increase of its temperature
$T_{p}$).${}^{9}$

Let $\nu_{pe}(q)$ be the characteristic phonon relaxation frequency 
owing to interaction with electrons. Gurevich at al.${}^{9}$ showed that
the value of $\nu_{pe}$ decreases rapidly for phonon wavevector $q \ge
2\bar p$, tending ro zero, where $\bar p$ is the average electron
momentum, namely $\bar p \sim \sqrt{2mT_{e}}$ for the nondegenerated
electron gas, and $\bar p \sim p_{F}$ (Fermi momentum) for the case of
degeneracy. This mean that electrons are scattered by phonones having
momentum values smaller than the average electron momentum ($q \le 
2\bar p$). Those phonones are knowen as long-wave (LW), in contrast to
short-wave (SW) phonones, whose momentum satisfy $q \ge 2\bar p$.
The degree of nonequilibrium of the LW (and hence SW) phonones is 
determinated by the relationship of the phonon-phonon collision
frequency $\nu_{pp}$ and $\nu_{pe}$:\\
i) Strong phonon-phonon interaction\\
This case satisfies the following inequality $\nu_{pp}\gg\nu_{pe}$. This
situation means that the phonon-phonon collisions are more frequent than
phonon-electron collisions and more efficient in terms of energy 
relaxation than energy transfer from the phonon to the electron system. 
As a result, the Planck distribution function describes the phonon 
system with its own temperature $T_{p}$ different from $T_{e}$. Since 
the phonon-phonon collision frequency depends on the phonon temperature,
the inequality $\nu_{pp}\gg\nu_{pe}$ holds for phonon temperature $T_{p}
> 50^{\circ}$K for $n$-type Ge and GaAs semiconductors with the charge 
density concentration $n\propto 10^{14}\textrm{ cm}^{3}$.${}^{9}$

ii) Strong phonon-electron interaction\\
Now, consider the situation represented by the inequality 
$\nu_{pe}\gg\nu_{pp}$. This is the case when phonon-phonon collisions 
alone cannot bring the phonon system to an internal equilibrium, and the
description becomes more complicated. If $T_{p}$ is the characteristic 
phonon temperature, than the momentum $T_{p}/s$ ($s$ is the sound 
velosity) sets a limitto the phase spacevolume occupied by phonons 
(with $q > T_{p} s$ the number of phonons is exponentially low). The 
number of LW and SW phonons depend on the relation between $2 \bar p$ 
and $T_{p}/s$. If $T_{p}/s \gg 2\bar p$ is true i.e., LW phonones occupy
a much smaller phase volume that SW phonones, then the subsystem of SW 
phonones has enough time to redistribute the energy recived from LW 
phonones between its constituent quasiparticles. As a result, the 
distribution function of SW phonones becones Plankian, with some 
temperature $T_{p}^{SW}$. Then the electron-LW phonon interraction relax
their energy more efficiently than the phonon subsystem and the LW 
phonons emitted by electrons of temperature $T_{e}$ are characterized by
the same temperature $T_{e} = T_{p}^{SW}$. In this situation we also 
have two different subsystems; one corresponds to th SW phonons with 
Temperature $T_{p}^{SW}$ and the other one corresponds to electrons and 
LW phonons with a characteristic temperature $T_{e} = T_{p}^{LW} \ne 
T_{p}^{SW}$. If $T_{p}/s \ll 2\bar p$ holds, then all phonons interact 
efficiently with electrons, and hence the phonon system cannot be 
subdivided into LW and SW phonons. In this case, the phonons emitted by 
electrons of temperature $T_{e}$ are characterized by the same 
temperature $T_{e} = T_{p}$ and all the phonons show $\nu_{pe} \gg 
\nu_{pp}$ for $T_{p} < 50^{\circ}$K.

At this point its important to mention that thermal waves phenomenon is 
present in the same only for modulated frequency of the incident light 
$\omega$ (chopper frequency) of the same order as the frequency of the 
relaxation energy between the quasiparticle systems 
$\nu_{\varepsilon}$.${}^{8}$ In the limit $\omega\gg\nu_{\varepsilon}$ 
the system cannot respond to this external perturbation; therefore the 
dynamic part of the heat flux is negligible as compared with the static 
part and the transfered heat is only static. For 
$\omega\ll\nu_{\varepsilon}$, the variation of the electron and phonon 
temperature oscillates with the same modulated frequency of the incident
light.${}^{10}$ For the case when $\nu_{pp}\gg\nu_{pe}$ then 
$\nu_{\varepsilon}\sim\nu_{pe}$ otherwise ($\nu_{pp}\ll\nu_{pe}$) 
$\nu_{\varepsilon}\sim\nu_{pp}$.${}^{9}$ Typical values of 
$\nu_{\varepsilon}$ are in the range of $10^{8}$--$10^{10}$ 
sec${}^{-1}$.${}^{6}$

Recently,$^{10}$ there has been some interest in studying thermal
characterization of layered systems using the photoacoustic effect.  In
particular, from the theoretical point of view, thermal diffusion of one
and two layer system has been investigated using the heat-diffusion
equation in the approximation when both the electron and phonon
temperature distribution are equal.  This approximation is only valid in
the limit of infinite electron-phonon energy interaction in which the
size of the sample is grater than the cooling lenght.${}^{7}$  The 
effective diffusivity and thermal conductivity of the two layer were
obtained assuming continuity of heat flux, instead of the temperature 
distribution, at each interface and taking into account the physical
solution of the dynamical part of the temperature fluctuation in each layer.

In this work we further extend the earlier model for thin film 
semiconductors by taking into account the so far neglected important
features the behavior of electtrons and phonons under time varying
excitation namely, the electron and phonon temperature distribution
function.  In Part~II the electron and phonon temperature distribution
for semiconductors is calculated by solving the coupled heat-diffusion
equations, for both of quasiparticles systems.  In Section~III a
discussion of our results is presented, and a comparison to the
predictions of the existing theories is also made.  Finally, in
Section~IV we present our conclusions.

\section{Formulation}
It is well known the heat transport in solids is carried out by various
quasiparticles (electrons, holes, phonons, magnons, plasmons, etc.). 
Frequently the interactions between these quasiparticles are such that
each of these subsystems can have its own temperature and the physical
conditions at the boundary of the sample.  We restrict our analysis to
the case of monopolar semiconductors under the conditions of strong 
phonon-phonon interaction asdiscussed previously.${}^{11}$ Therefore 
steady state heat conduction can be described by the following system of
equations$^9$.
\begin{equation}
\operatorname{div}\mathbf{Q}_{e} = 
-P_{ep}(T_{e} -T_{p}),\qquad 
 \operatorname{div}\mathbf{Q}_{p} =
 P_{pe}(T_{e} - T_{p}). 
\end{equation}
The term $P_{ep}(T_{e} - T_{p})$ describes the transfer of heat between 
electrons and phonons.  Here $P_{ep}$ is a parameter proportional to the
energy frequency between electron and phonon systems 
($P_{ep}=P_{pe} = P \sim n\nu_{\varepsilon}$) and the heat flux of 
electron $\mathbf{Q}_{e}$ and phonon $\mathbf{Q}_{p}$ subsystems are 
described by the usual relationships:
\begin{equation}
\mathbf{Q}_{e}= -\varkappa_{e} \operatorname{grad}T_{e},\qquad 
\mathbf{Q}_{p}= -\varkappa_{p} \operatorname{grad}T_{p} 
\end{equation}
where $\varkappa_{e}$ ($\varkappa_{p})$ is the electron (phonon) thermal 
conductivity.  So far, Eqs.~(1) include only the static contribution of 
the heat transport, i.e. the heat flux is independent of time. However,
in the photothermal experiments, the incident radiation is modulated in
time by the chooper, and in this case it is necessary to consider the
dynamic contribution to the heat transport in the electron and phonon
systems.  Let us assume the one dimensional model for the heat flux. 
Assuming that the sample is optically opaque to the incident light
(i.e., all the incident light is absorbed at the surface), the electron
and phonon temperature distribution function are the solutions of
\begin{equation}
\frac{\partial^{2}T_{e}(z,t)}{\partial z^{2}}
- k^{2}_{e}
\left[
T_{e}(z,t) - T_{p}(z,t)
\right]
=
\frac{1}{\alpha_{e}}
\frac{\partial T_{e}(z,t)}{\partial t},
\end{equation}
\[
\frac{\partial^{2}T_{p}(z,t)}{\partial z^{2}}
+ k^{2}_{p}
\left[
T_{e}(z,t)-T_{p}(z,t)
\right]
=
\frac{1}{\alpha_{p}}
\frac{\partial T_{p}(z,t)}{\partial t} ,
\]
where $k^{2}_{e,p}=\frac{P}{\varkappa_{e,p}}$, and the diffusivity
for each system is given as 
$\alpha_{e,p}=\varkappa_{e,p}/{(\rho c)}_{e,p}$ 
and $\rho_{e}$ ($\rho_{p}$), $c_{e}$ ($c_{p}$) is the electron (phonon) 
density and specific heat respectively.

The temperature fluctuation $T_{e,p}(z,t)$ should be supplemented by 
boundary conditions at the surface of the semiconductor $(z=0)$.  In the
photothermal experiment, the most common mechanism to produce thermal
waves is the absorption by the sample of an intensity modulated light
beam with frequency modulation $\omega\le\nu_{\varepsilon}.$  It is 
clear that when the intensity of the radiation is fixed, the
light-into-heat convension at the surface of the sample can be written
in general as
\begin{equation}
\left.
Q_{e,p}(z,t)
\right|_{z=0} =
Q_{e,p}+\Delta Q_{e,p}e^{i\omega t}
\end{equation}
where $Q_{e,p}$ is proportional to the intensity of high frequency 
light and the other term represents the modulation of this light.

The general solution of the coupled heat-diffusion equation for the 
electron and phonon system can be written as
\begin{equation}
T_{e,p} = 
A + Bz 
\pm\frac{k^{2}_{e,p}}{k^{2}}Ce^{-kz} 
+ \theta_{e,p}(z)e^{i\omega t}  
\end{equation}
where $k^{2} = k^{2}_{e} + k^{2}_{p}$ represents the 
inverse of cooling lenght$^7$ and $\theta_{e,p}(z)$ satisfies a similar
set of Eqs. (3) but instead of the term
$(1/\alpha_{e,p})\partial T_{e,p}/\partial t$ we substitute
$(i\omega /\alpha_{e,p})T_{e,p}$, and the solution is given by
\begin{equation}
\theta_e(z)=Fe^{- \sigma z}
\end{equation}
\[
\theta_p(z)=Ge^{- \sigma z}.
\]
Here $F$ and $G$ are related each other by the following relationship
\begin{equation}
G = - \frac{\sigma^{2} - \sigma^{2}_{e}}{k^{2}_{e}}F
\end{equation}
and the values of $\sigma$ are given by
\begin{equation}
\sigma^2_{1,2} = 
\frac{1}{2}
\left(
\sigma^{2}_{e}+\sigma^{2}_{p}
\right)
\pm\frac{1}{2}
{\left[
{(\sigma^{2}_{e} - \sigma^{2}_{p})}^{2} 
+ 4 k^{2}_{e} k^{2}_{p}
\right]}^{1/2}
\end{equation}
Eq.~(8) represents the condition for non-trivial solutions of the 
coupled elec\-tron-phonon differential equations and $\sigma^2_{e,p}$ 
are given by
\begin{equation}
\sigma^{2}_{e,p} =
\frac{i\omega}{\alpha_{e,p}} + k^{2}_{e,p} .
\end{equation}
It is worth mentioning that the increasing exponential term 
$\exp{kz}$ and $\exp{\sigma_{1,2}z}$ which are solution of the 
static and the dynamical part of heat-diffusion equations respectively 
have not been considered because they do not represent a physical 
solution (heat flux cannot be reflected).

Using the boundary conditions at the surface of the sample given by 
Eq.~(4) the electron and phonon temperatures are given by
\[
T_{e} = A + Bz + \frac{k_{e}^{2}}{k^{2}} C e^{-kz} + 
e^{i\omega t} 
\left[
F_{1}e^{-\sigma_{1}z} + F_{2} e^{-\sigma_{2}z}
\right],
\]
\begin{equation}
T_{p} = A + Bz + \frac{k_{p}^{2}}{k^{2}} C e^{-kz} + 
e^{i\omega t} 
\left[
G_{1}e^{-\sigma_{1}z} + G_{2} e^{-\sigma_{2}z}
\right],
\end{equation}
where the constants $B$, $C$, $F_{1,2}$, $G_{1,2}$ can be written as
\begin{equation}
B = -\frac{1}{k^2}
\left(
\frac{Q_e}{\varkappa_e}k^2_p + \frac{Q_p}{\varkappa_p}k^2_e
\right)
\tag{\theequation{a}}
\end{equation}
\begin{equation}
C = 
\frac{1}{k}
\left(
\frac{Q_e}{\varkappa_e}-\frac{Q_p}{\varkappa_p}
\right)
\tag{\theequation{b}}
\end{equation}
\begin{equation}
F_1 = 
\frac{1}{\sigma_1(\sigma^2_2-\sigma^2_1)}
\left[
\frac{\Delta Q_{p}}{\varkappa_p} k^2_e +
\frac{\Delta Q_{e}}{\varkappa_e}
(\sigma^2_2-\sigma^2_e)
\right]
\tag{\theequation{c}}
\end{equation}
\begin{equation}
F_2 = - 
\frac{1}{\sigma_2(\sigma^2_2-\sigma^2_1)}
\left[
\frac{\Delta Q_p}{\varkappa_p} k_{e}^{2} +
\frac{\Delta Q_e}{\varkappa_e} (\sigma^2_1-\sigma^2_e)
\right]
\tag{\theequation{d}}
\end{equation}
\begin{equation}
G_{1,2} =
- \frac{\sigma_{1,2}^{2} - \sigma_{e}^{2}}{k_{e}^{2}} F_{1,2}
\tag{\theequation{e}}
\end{equation}\stepcounter{equation}
Here $A$ is a constant which cannot be determinated from these boundary
conditions and it is not important in obtaining the physical results.

Once we know the electron and phonon temperature distributions in the 
sample, we can calculated the responce of the surrounding medium due 
to thephotothermal heating of the sample using one of the several 
alternative detection techniques mentioned before.

\section{Special Cases}
We now turn to a discussion on the results obtained so far and compare
them with previous theories on thermal waves.  We shall first consider a
nondegenerate semiconductors. In this case, the typical ratio of the 
heat conductivity of electrons to phonon satisfies
$\varkappa_e/\varkappa_p\sim 10^{-3}$ then $k_e\gg k_p$.
Under these circunstances, after simplying the expression for $T_e$, 
$T_p$ and $\sigma_{1,2}$ can be written as
\begin{equation}
T_e(z,t) = 
A + Bz + Ce^{-kz}+e^{i\omega t}
\left(F_1 e^{-\sigma_{1}z} + F_2e^{-\sigma_{2}z}\right)
\tag{\theequation{a}}
\end{equation}
\begin{equation}
T_p(z,t) = 
A + Bz - \frac{k^2_p}{k^2_e}Ce^{-kz}+G_2e^{i\omega t-\sigma_2z}
\tag{\theequation{b}}
\end{equation}\stepcounter{equation}
where $G_1\approx 0$, $\sigma_1=\sigma_e$ and 
$G_{2} = \left[1 - 
\frac{i\omega}{k_{e}^{2}}\left(\frac{1}{\alpha_{p}} - 
\frac{1}{\alpha_{e}}\right)\right] F_{2}$, 
$\sigma_2=i\omega /\alpha_p$. It is important to note that the
contribution of the electron diffusivity dissapears in the phase of the
dynamical part of the phonon temperature.

Because the main source of the photothermal signal arieses from the 
periodic heat flow from the semiconductor, the periodic diffusion 
process produces a periodic temperature variation in the semiconductor 
given by the sinusoidal (AC) component of Eqs.~(12).

Information about electron and phonon parameters can be obtained from 
photothermal experiments depending upon the decay lenght
$s_{1,2}=\Re\sigma_{1,2}$ of the AC component of the electron and 
phonon temperature at $z=d$ (where $d$ is the thickness of the sample).

As can be seen in Fig.~1a, if $|F_{1}| \gg |G_{2}| \ge |F_{2}|$ and 
$s_{1} \ll s_{2}$ the time dependent component of the phonon temperature
in the semiconductor attenuates rapidly to zero with increasing distance
from the surface of the solid as compared with component of the electron
temperature, and the electron thermal wave carried out all the 
informationabout electron and phonon parameters through the coefficient 
$F_{1}$. However if besides that $\frac{\Delta 
Q_{p}}{\varkappa_{p}}\ll\frac{\Delta Q_{e}}{\varkappa_{e}}$ electron 
thermal wave provide us only information about thermal parameters of the
electron system. Now if 
$\frac{\Delta Q_{p}}{\varkappa_{p}}\gg
\frac{\Delta Q_{e}}{\varkappa_{e}}$
information about the phonon system is only obtained from the 
photothermal experiments.

On the other hand if $|F_{1}|\gg|G_{2}|\ge|F_{2}|$ and $s_{1}\gg s_{2}$,
in this case the time dependent component of the electron temperature is
fully damped out as shown in Fig.~1b. Information about the electron or 
phonon parameters depend on the relation between $d$ and $s_{1}^{-1}$ 
and how the incident energy is distributed in the electron and phonon 
systems:\\
(1) $d < s_{1}^{-1}$ and $\frac{\Delta Q_{e}}{\varkappa_{e}}\gg
\frac{\Delta Q_{p}}{\varkappa_{p}}$, Eqs.~(11c) and (12a) tell us that 
thermal wave asociated with the electron system only give information 
about the electron thermal parameters.\\
(2) $d > s_{1}^{-1}$ and $\frac{\Delta Q_{e}}{\varkappa_{e}}\ll
\frac{\Delta Q_{p}}{\varkappa_{p}}$, in this case only a phonon thermal 
wave can be detected at $z=d$ and the photothermal experiments give the 
thermal phonon parameters of the sample.

Similar behavoir of photothermal waves in the semiconductors can be 
obtained if $|G_{2}|\gg|F_{2}|,|F_{1}|$ and $s_{2}\ll s_{1}$ and 
information about phonon thermal parameters is obtained from the 
photothermal experiments if in additionthe inequality 
$\frac{\Delta Q_{p}}{\varkappa_{p}}\gg
\frac{\Delta Q_{e}}{\varkappa_{e}}$ 
holds. If $|G_{2}|\gg|F_{2}|,|F_{1}|$ and $s_{2}\gg s_{1}$, hence the 
photothermal experiments only detect the thermal parameters of the 
phonon system when $d\ll s_{2}^{-1}$ and 
$\frac{\Delta Q_{p}}{\varkappa_{p}}\gg
\frac{\Delta Q_{e}}{\varkappa_{e}}$ 
or thermal parameters of the electron system if $d\gg s_{2}^{-1}$ and
$\frac{\Delta Q_{e}}{\varkappa_{e}}\gg
\frac{\Delta Q_{p}}{\varkappa_{p}}$  as 
show in Figs.~ 2a and 2b.

In the case $k_{e,p}\to 0$ (electron-phonon interaction vanishes), 
it follows from Eqs.~(11) that the electron and phonon temperatures 
reduce to
\begin{equation}
T_{e,p} =
A_{0} - \frac{Q_{e,p}}{\varkappa_{e,p}} z +
\frac{\Delta Q_{e,p}}{2\varkappa_{e,p}}
{\left(\frac{2\alpha_{e,p}}{\omega}\right)}^{1/2}(1 - i) 
e^{i\omega t - \sigma_{e,p} z } 
\end{equation}
whete the constant $A$ has been chosen such it exactly cancels out the 
divirgent term from the Eq.~(11b), i.e. for $P\to 0$ we have 
non-interacting systems of quasiparticles and heat transport is carried 
out by the electrons and phonon independently through the semiconductor.

We now consider, the situation represented by the strong coupling
between electrons and phonons i.e. $P\to\infty$.  In this case
the electron-phonon energy interaction is very efficient in terms of
energy relaxation than energy transfer in the electron or phonon system. 
Here the solutions of Eq. (8) are given by
\[
\sigma^2_1=k^2 + \frac{i\omega}{k^{2}}
\left(
\frac{k_{e}^{2}}{\alpha_{e}} +
\frac{k_{p}^{2}}{\alpha_{p}}
\right)
\]
\begin{equation}
\sigma^2_2 =
\frac{i\omega}{k^{2}}
\left(
\frac{k^2_p}{\alpha_e}+\frac{k^2_e}{\alpha_p}
\right)
\end{equation}
and $G_2\approx F_2$, 
$B = - 
\frac{1}{\varkappa_{e} + \varkappa_{p}}\left(Q_{e} + Q_{p}\right)$
therefore, in this limit both temperatures are the same
\begin{equation}
T=T_e=T_p=A+Bz+F_2e^{i\omega t-\sigma_2z}
\end{equation}
Here the amplitude and decay lenght of temperature fluctuation in the 
sample depend in a nontrivial way on the diffusivity, heat conductivity
of electron and phonon system and energy frequency of electron-phonon
interaction.  Equation (15) can be obtained from Eqs. (3) in the limit
when $P\to\infty$.  In this situation the electron and phonon 
temperature are the same, the product $P(T_e-T_p)$ approaches to 
unknown finite value and the set of differential equations for
$P(T_e-T_p)$ and $T_e=T_p=T$ reduce to one,
\begin{equation}
\frac{\partial^2T}{\partial z^2} =
\frac{1}{\alpha}
\frac{\partial T}{\partial t} 
\end{equation}
with
\begin{equation}
\frac{1}{\alpha}=\frac{1}{k^2}
\left(
\frac{k^2_p}{\alpha_e}+
\frac{k^2_e}{\alpha_p}
\right)
\end{equation}
where the solution is similar to Eq. (15) with the appropiate boundary 
condition at the surface of the semiconductor given by
\begin{equation}
\left.Q(z,t)\right|_{z=0} =
\left. Q_{e}(z, t) + Q_{p}(z, t)\right|_{z=0} =
Q_{e} + Q_{p} + \left(\Delta Q_{e} + \Delta Q_{p}\right)e^{i\omega t}
\end{equation}

At this point, the authors beleive that is important to remark the 
difference between thermal waves and electromagnetic waves (for 
example).  In the absence of rigorous theoretical guidance from first 
principles, several workers find it necessary to introduce arbitrary 
algebraic factor into their calculations in oder to get desirable fit to
the data. As a consequence the controversial analogy between termal 
waves and electromagnetic waves arises.

It is our contention that the main differences are the following:\\
i) As pointed out by Mandelis${}^{12}$ electromagnetic waves satisfy a 
hyperbolic differential equation while in the case for thermal waves 
satisfy a parabaloic differential equation. In other words heat 
conduction in solid is a diffusive process, quite different to 
electromagnetic wave propagation.\\
ii) It is well known from standard electromagnetic books that the 
tangential component of the electric field at the interface between two 
different media must to be continuous. In addition there are a reflected
and transmited electromagnetic waves through the interface. However for 
the thermal waves there is not such similarity i.e., the electron or 
phonon temperature distributions in solid is not continuous in general 
at the interface of two layer system (see Ref.~10) only heat flux 
(energy conservation) is continuous at the boundary between two thermal 
different media.\\
iii) It is also well known that a damped electromagnetic wave in a 
medium is of the form $e^{i(\omega t -kx) - \lambda x}$ where in 
general the decaying length staisfies the following inequality 
$\lambda^{-1}\ge k^{-1}$ where $k$ is the wave vector associated with 
the electric field. as mentioned before, information about electron and 
phonon thermal parameters can be obtained from photothermal experiments.
In particular when the electron-phonon interaction is strong such that 
$k_{e}^{2}\gg\omega/\alpha_{e}$, in this limit the decay length 
$(\Re\sigma_{1})^{-1}$ is smaller than the wave length 
$(\Im\sigma_{1})^{-1}$ of the thermal wave associated with the 
electron system. This behavoir is impossible to get from the propagation
of electromagnetic waves. This three points show that the analogy 
between thermal and electromagnetic waves is not correct.

\section{Conclusions}
A theoretical analysis of thermal waves in thin films semiconductors has
been studied.  Using the appropriate boundary conditions, we obtain the
electron and phonon temperature distribution taking into account the
physical solution for the static and dynamical parts of the
heat-diffusion equation.  For typical parameters of the heat 
conductivity el electrons and phonon in semiconductors ($k_e\gg k_p$), 
it is possible to obtain information about the physical parameteres
describing the diffusivity and electron-phonon interaction in
photoacoustic experiments.  It is shown, Eq. (12), that the photothermal
signal is ultimately governed by the electron or phonon system depending
on the relationship of the amplitude and the decay lenghts of the 
temperature fluctuation in each system and the sample size.  We also
showed that it is possible that decaying parameter of the thermal wave 
be smaller than the wavelength.  However, it is well known that from
Maxwell equations this situation is forbiden for electromagnetic waves.

We have also derived exact solutions for electron and phonon 
temperature in semiconductors in the limit of weak and strong
electron-phonon interactions. The above findings tell us that thermal
waves in semiconductors can propagate independent each other or 
propagate as a thermal wave in the sample with an effective diffusivity
incluiding the contribution of the thermal parameters of the electron 
and phonon system.

\section*{Asknowledgments}

This work is partially support by Consejo Nacional de Ciencia y 
Tecnolog\'{\i}a (CONACyT), M\'{e}xico.

\end{document}